\documentclass[a4paper, oneside, twocolumn, notitlepage, 10pt]{extarticle_ecoc}
\usepackage{ecoc}
\pdfoutput=1
\usepackage{tikz,pgfplots}
\usepackage{graphicx}
\usepackage{placeins}
\usepackage{mfirstuc}
\usepackage{mathtools}
\usepackage{wrapfig}
\usepackage{pdfpages}
\usepackage{diagbox}
\usepackage{makecell}
\usepackage{amssymb}
\usepackage{bbm}
\usepackage{acronym}
\pgfplotsset{compat=1.16}
\binoppenalty=\maxdimen
\relpenalty=\maxdimen
\usetikzlibrary{backgrounds,shapes,arrows,positioning}
\usetikzlibrary{shapes,snakes}
\usetikzlibrary{fit,calc} %

\usetikzlibrary{fit}
\usetikzlibrary{decorations.pathreplacing}
\definecolor{KITpalegreen}{RGB}{130,190,60}
\definecolor{KITcyanblue}{RGB}{80,170,230}
\definecolor{KITorange}{rgb}{.87,.60,.10}
\definecolor{tab104}{RGB}{152,78,163}
\definecolor{colorDRSD}{RGB}{255,128,0}
\definecolor{colorideal}{RGB}{224,243,248}
\definecolor{colorDRSDdeter}{RGB}{168,129,188}
\definecolor{colorDRSDplus20}{RGB}{255,102,178}
\definecolor{colorBEE-PC}{RGB}{96,96,96}
\definecolor{colorAD}{RGB}{0,102,51}
\definecolor{orange1}{RGB}{254,196,79}%
\definecolor{orange2}{RGB}{254,153,41}
\definecolor{orange3}{RGB}{217,95,14}
\definecolor{orange4}{RGB}{153,52,4}

\definecolor{pink1}{RGB}{250,159,181}%
\definecolor{pink2}{RGB}{247,104,161}
\definecolor{pink3}{RGB}{197,27,138}
\definecolor{pink4}{RGB}{122,1,119}

\definecolor{mygreen}{rgb}{0.69, 0.87, 0.541}%
\definecolor{myblue}{rgb}{0,0.4470,0.7410}
\definecolor{myblack}{rgb}{0.2,0.2,0.2}
\definecolor{proposed5}{RGB}{188,189,220}
\definecolor{proposed4}{RGB}{158,154,200}
\definecolor{proposed1}{RGB}{128,125,186}
\definecolor{proposed2}{RGB}{106,81,163}
\definecolor{proposed3}{RGB}{84,39,143}
\definecolor{proposedfull}{RGB}{63,0,125}

\usepackage{bm}
\renewcommand{\vec}[1]{\bm{#1}}
\newcommand{\textsub}[1]{\textnormal{#1}}
\newcommand{\CW}{\mathcal{C}}

\newcommand{\dmin}{d_{\textnormal{min}}}
\newcommand{\ddesign}{d_{\textup{des}}}

\newcommand{\que}{\mathord{?}}

\usepackage{bm}
\renewcommand{\vec}[1]{\bm{#1}}
\DeclareMathOperator{\DF}{\mathsf{D}}
\newcommand{\DFC}{\DF_{\textsub{C}}}
\newcommand{\BDD}{\textnormal{BDD}}

\newcommand{\Eb}{E_{\textnormal{b}}}
\newcommand{\No}{N_{\textnormal{0}}}

\newcommand{\Ta}{T_{\mathrm{a}}}
\newcommand{\na}{n_{\mathrm{a}}}
\newcommand{\nee}{n_{\mathrm{e}}}

\newcommand{\Te}{T_{\mathrm{e}}}

\newcommand{\pone}{\vec{p}^{(\textnormal{1})}}
\newcommand{\ptwo}{\vec{p}^{(\textnormal{2})}}
\newcommand{\wone}{\vec{w}^{(\textnormal{1})}}
\newcommand{\wtwo}{\vec{w}^{(\textnormal{2})}}
\newcommand{\poneJ}{\vec{p}^{(\textnormal{1})}_{J}}
\newcommand{\ptwoJ}{\vec{p}^{(\textnormal{2})}_{J}}

\newcommand{\woneJ}{\vec{w}^{(\textnormal{1})}_{J}}
\newcommand{\wtwoJ}{\vec{w}^{(\textnormal{2})}_{J}}

\tikzset{mark options={mark size=3, line width=1pt}}
\definecolor{colorCapacityGainAWGN}{RGB}{152,78,163}
\definecolor{myblue}{rgb}{0,0.4470,0.7410}

\acrodef{MC}{miscorrection}
\acrodef{HDD}{hard decision decoding}
\acrodef{HD}{hard decision}
\acrodef{SDD}{soft decision decoding}
\acrodef{TPD}{turbo product decoding}
\acrodef{BSC}{binary symmetric channel}
\acrodef{iBDD}{iterative bounded-distance decoding}
\acrodef{BDD}{bounded distance decoding}
\acrodef{SA-HDD}{soft-aided \ac{HDD}}
\acrodef{PC}{product code}
\acrodef{AD}{anchor decoding}
\acrodef{HRB}{highly reliable bit}
\acrodef{EaE}{error-and-erasure}
\acrodef{EaED}{error-and-erasure decoder}
\acrodef{SABM}{soft-aided bit marking}
\acrodef{iEaED}{iterative error-and-erasure decoding}
\acrodef{BI-AWGN}{binary input additive white Gaussian noise}
\acrodef{BCH}{Bose--Chaudhuri--Hocquenghem}
\acrodef{DRS}{dynamic reliability score}
\acrodef{DRSD}{\ac{DRS} decoder}
\acrodef{BER}{bit error rate}
\acrodef{SABM-SR}{SABM with scaled reliabilities}
\acrodef{NCG}{net coding gain}
\acrodef{GMD}{generalized minimal distance}
\acrodef{GMDD}{generalized minimal distance decoder}
\acrodef{RS}{Reed--Solomon}
\acrodef{DE}{density evolution}
\acrodef{LLR}{log-likelihood ratio}
\acrodef{BPSK}{binary phase shift keying} 
\acrodef{SNR}{signal-to-noise ratio} 
\acrodef{BEE-PC}{binary message passing based on \ac{EaE} decoding for \acp{PC}}
\acrodef{EMP}{extrinsic message passing}
\acrodef{SISO}{soft input soft output}
\acrodef{HIHO}{hard input hard output}

\begin{document}
	\selectlanguage{english}    %

	\title{Improved Soft-aided Decoding of Product Codes with Adaptive Performance-Complexity Trade-off}%

	\author{
		Sisi Miao, Lukas Rapp, and Laurent Schmalen
	}
	
	\maketitle                  %

	\begin{strip}
		\begin{author_descr}
			
			Communications Engineering Lab (CEL), Karlsruhe Institute of Technology (KIT), Karlsruhe (Germany)
			\textcolor{blue}{\uline{sisi.miao@kit.edu}}
		\end{author_descr}
	\end{strip}
	
	\setstretch{1.1}
	\newcommand\extrafootertext[1]{%
		\bgroup
		\renewcommand\thefootnote{\fnsymbol{footnote}}%
		\renewcommand\thempfootnote{\fnsymbol{mpfootnote}}%
		\footnotetext[0]{#1}%
		\egroup
	}
	\renewcommand\footnotemark{}
	
	\begin{strip}
		\begin{ecoc_abstract}
			We propose an improved soft-aided decoding scheme for product codes that approaches the decoding performance of conventional soft-decision TPD with only a 0.2$\;$dB gap while keeping the complexity and internal decoder data flow similarly low as in hard decision decoders. \textcopyright2022 The Author(s)\vspace{-0.5\baselineskip}
		\end{ecoc_abstract}
	\end{strip}

    \section{Introduction}
	\extrafootertext{This work has received funding from the European Research Council (ERC) under the European Union's Horizon 2020 research and innovation programme (grant agreement No. 101001899).}
	\Acp{PC}~\cite{elias1955coding} are powerful code construction for high rate fiber-optical communication. Conventionally, \acp{PC} are decoded either with \ac{SDD}, with one popular and powerful instance being \ac{TPD}~\cite{pyndiah1998near}, or with \ac{HDD}, e.g. with the ubiquitous \ac{iBDD} decoder. \ac{SDD} with TPD provides excellent error correcting performance but entails a high complexity and a large internal decoder data flow. Although the decoding performance of HDD with iBDD is not as satisfying as  the one of TPD, iBDD is still of great interest for high speed communication systems due to its low complexity. Many works have been devoted to seek complexity-performance trade-offs between SDD and HDD. Multiple approaches to reduce the decoding complexity of the TPD decoder have been presented, e.g., in~\cite{kaneko1994,Dave2001efficient,AlDweik2009hybrid,Chen2009Testpattern,AlDweik2018Ultralight,Wang2021Investigation}. However, its complexity is still significantly higher than the complexity of iBDD.
	A more recent direction of research uses a certain amount of soft information to aid HDD of PCs~\cite{sheikh2018iterative,sheikh2018low,sheikh2019binary,sheikh2019BMPGMDD,sheikh2021refined,liga2019novel,lei2019improved,sheikh2021novel}. We refer to this family of decoders as \emph{soft-aided} decoders. Such decoders do not require soft message passing and are based on HDD component decoders, thus the complexity is similarly low as in iBDD. However, a significant decoding performance gap is observed compared to TPD. A common issue among existing soft-aided decoders is that the soft channel information remains static and is not updated during iterative decoding, as in TPD. We addressed this issue in our recent works by introducing \acp{DRS}~\cite{miao2021improved,miao2022journal}. The resulting \ac{DRSD} provides the best decoding performance for PCs so far among soft-aided decoders. In this paper, we improve upon our previous work and propose a novel soft-aided row/column decoder that uses DRSs to further improve the error-correcting performance. The proposed decoder can realize a performance-complexity trade-off and approaches the TPD performance.
	
	\vspace{-0.5\baselineskip}
	\section{Product Codes and Decoding Model}	For illustrative purpose, the PCs considered in this paper consist of identical row/column component codes $\mathcal{C}[n,k,t]$, which are even-weight subcodes of primitive BCH codes with $n=2^{\nu}-1$ and $k=2^{\nu}-t\nu -2$. The PC has rate $r=k^2/n^2$. The even-weight subcode increases the minimal distance of the component code by one such that $\dmin\geq 2t+2 =: \ddesign$. We consider \ac{BPSK} modulation and assume that the codewords are transmitted over a \ac{BI-AWGN} channel. For any transmitted bit $x_i$, the channel output is $\tilde{y}_i=(-1)^{x_i}+n_i$,
	where $n_i$ is (real-valued) AWGN with noise variance $\sigma^2_n = (2r\Eb/\No)^{-1}$.
	
	In iterative decoding of PCs, the rows and columns of the PC block are alternately decoded with a component code decoder $\DFC$ until either the entire PC block is successfully decoded or the maximum number of iterations $L$ is reached. In HDD/iBDD decoders, $\DFC$ is a simple BDD. Decoding can be performed in the syndrome domain and the estimated internal decoder data flow of an iBDD decoder is two orders of magnitude smaller than the corresponding data flow in SDD of LDPC codes~\cite{smith2012staircase}. In SDD/TPD decoders, $\DFC$ is a list-based SD decoder that generates multiple candidate codewords. The decoding decision is determined based on the Euclidean distance between the candidate codewords and the real-valued row/columns vector. The computational overhead is due to the increased number of BDD steps and also a large number of float number arithmetic operations in the candidate codeword selection. As investigated in~\cite{Wang2021Investigation}, the latter causes twice the amount of power consumption as the former. After each decoding step, soft extrinsic messages are calculated and passed for every bit, which entails an internal decoder data flow orders of magnitude higher than the transmitted binary data itself. This motivated the research on soft-aided HDD schemes, where only simple arithmetic operations are required and no soft messages are passed.

	\vspace{-0.5\baselineskip}
	\section{Proposed Decoder}
	To enable a list-based component code decoder, we turn the \ac{BSC} channel into an \ac{EaE} channel. Received values which are close to hard decision threshold are marked as erasures to avoid introducing errors. For every received bit $\tilde{y}_i$, the ternary value $y_i$ is obtained by\vspace{-0.5\baselineskip}
	\begin{equation*}
	y_i = \begin{cases} \que~\textnormal{(erasure)} & \textnormal{if}\;|\tilde{y}_i|<T\\\textnormal{sign}(\tilde{y}_i)&\textnormal{if}\;|\tilde{y}_i|\geq T,\end{cases}\vspace{-0.5\baselineskip}
	\end{equation*}
	where $T$ is an optimizable threshold. This is based on the fact that the reliability of the received bits $y_i$ is proportional to the magnitude of $\tilde{y}_i$. We additionally use the \ac{DRS} introduced in~\cite{miao2021improved,miao2022journal} which coarsely represents the reliability and can be updated with hard messages. The \acp{DRS} are represented by 5-bit integers in the range $[0,31]$ (as we observe that further increasing the number of representation levels does not improve the performance significantly). We assign an initial DRS for each bit according to its magnitude $|\tilde{y}_i|$. The values of $|\tilde{y}_i|$ are sorted ascendingly and then evenly divided into $24$ groups (the last group may have fewer entries when $n^2$ does not divide $24$). The bits in the first group (lowest reliability) will have DRS 8 and the bits in the last group (highest reliability) will have DRS 31. Then we proceed with decoding the rows and columns with the proposed component code decoder $\DFC$.

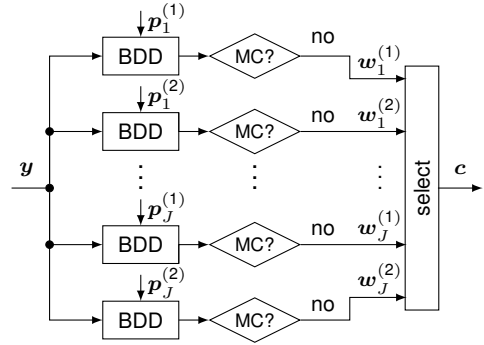
\begin{figure}[t!]
		\centering
		\tikzset{line/.style={-latex}} 
		\begin{tikzpicture}
    \node[draw=none] at (0,3) {\footnotesize$\vec{y}$};
    \node[draw,circle,fill,inner sep=1pt] at (0.3,2.75)(y) {};
    \node[draw,circle,fill,inner sep=1pt] at (0.3,3.5){};
    \node[draw,circle,fill,inner sep=1pt] at (0.3,2) {};
    \draw[] (-0.2,2.75) -- (y);
    \draw[] (1,4.25) rectangle ++(1,0.5);
    \node[draw=none] at (1.5,4.5)(BDD11) {\footnotesize$\BDD$};
    \node[draw=none] at (1.85,5) {\footnotesize$\pone_1$};
    \draw[line] (1.5,5.1)--(1.5,4.75);
    \draw[line] (y) |- (1,4.5);
    \draw[] (1,3.25) rectangle ++(1,0.5);
    \node[draw=none] at (1.85,4) {\footnotesize$\ptwo_1$};
    \node[draw=none] at (1.5,3.5)(BDD12) {\footnotesize$\BDD$};
    \draw[line] (1.5,4.1)--(1.5,3.75);
    \draw[line] (y) |- (1,3.5);
    \node[draw=none] at (1.5,3) {$\vdots$};
    
    \draw[] (1,1.75) rectangle ++(1,0.5);
    \node[draw=none] at (1.5,2)(BDDj1) {\footnotesize$\BDD$};
    \node[draw=none] at (1.85,2.5) {\footnotesize$\poneJ$};
    \draw[line] (1.5,2.6)--(1.5,2.25);
    \draw[line] (y) |- (1,2);
    \draw[] (1,0.75) rectangle ++(1,0.5);
    \node[draw=none] at (1.85,1.5) {\footnotesize$\ptwoJ$};
    \node[draw=none] at (1.5,1)(BDDj2) {\footnotesize$\BDD$};
    \draw[line] (1.5,1.6)--(1.5,1.25);
    \draw[line] (y) |- (1,1);

    \node [draw, diamond,aspect=2, inner sep=0.05cm, anchor=center](MC11)at (3,4.5) {\scriptsize MC?};
    \node[draw=none] at (3.9,4.75) {\footnotesize no};
    \node [draw, diamond,aspect=2, inner sep=0.05cm, anchor=center](MC12)at (3,3.5) {\scriptsize MC?};
    \node[draw=none] at (3.9,3.7) {\footnotesize no};
    \node [draw, diamond,aspect=2, inner sep=0.05cm, anchor=center](MCj1)at (3,2) {\scriptsize MC?};
    \node[draw=none] at (3.9,2.2) {\footnotesize no};
    \node [draw, diamond,aspect=2, inner sep=0.05cm, anchor=center](MCj2)at (3,1) {\scriptsize MC?};
    \node[draw=none] at (3.9,1.2) {\footnotesize no};
    \node[draw=none] at (3,3) {$\vdots$};
    \draw[line] (2,4.5) |- (MC11);
    \draw[line] (2,3.75) |- (MC12);
    \draw[line] (2,1.75) |- (MCj1);
    \draw[line] (2,1) |- (MCj2);

    \node[draw,rectangle,rotate=90,text width=3cm, minimum height=0.4cm,align=center] at (5.2,2.75)(w) {\footnotesize select};
    \draw[line] (MC11) -| (4.25,4.2)--(5,4.2);
    \draw[line] (MC12) -| (4.15,3.5) -- (5,3.5);
    \draw[line] (MCj1) -| (4.15,2) -- (5,2);
    \draw[line] (MCj2) -| (4.25,1.3)--(5,1.3);
    
     \node[draw=none] at (4.65,4.45) {\footnotesize $\wone_1$};
     \node[draw=none] at (4.65,3.75) {\footnotesize $\wtwo_1$};
     \node[draw=none] at (4.65,3) {\footnotesize $\vdots$};
     \node[draw=none] at (4.65,2.25) {\footnotesize $\woneJ$};
     \node[draw=none] at (4.65,1.55) {\footnotesize $\wtwoJ$};

     \node[draw=none] at (5.7,3) {\footnotesize$\vec{c}$};
     \draw[line] (5.41,2.75) -- (6,2.75);
   
		\end{tikzpicture}
		\caption{Block diagram of the proposed row/column decoder when $E\geq 1$.}
		\vspace{-0.5\baselineskip}
		\label{fig:Dc}
	\end{figure}

	The proposed decoder $\DFC$ is a list-based decoder and the decoding decision is determined based on Hamming distances and simple integer computations. The list size is associated with $\mathcal{J}$, which can be configured to achieve a performance-complexity trade-off. For every received row/column $\vec{y}$, let $E$ denote the number of erasures in $\vec{y}$. If $E>\ddesign$, we do not decode and proceed to the next row/column as too many erasures cannot be handled by the decoder. If $0\leq E\leq \ddesign$, we generate a set of candidate codewords. If $E=0$, a usual BDD step is performed and only one candidate codeword is obtained. If $E\geq 1$, the block diagram of the proposed $\DFC$ is shown in Fig.~\ref{fig:Dc}. We define a \emph{filling pattern} for the erasures consisting of a pair of binary complementary \emph{random} vectors of length $E$ $(\pone, \ptwo)$ where $\pone+\ptwo=(1,1,\ldots,1)$. Let $J=\min(E-1,\mathcal{J})$ as there are only $2^{E-1}$ distinct filling patterns for $E$ erasures. We generate $J$ pairs of test patterns by replacing the erasures in $\vec{y}$ with filling patterns $(\pone_{j}, \ptwo_{j})$, where $j\in \{1,2,\ldots,J\}$. The filling patterns are chosen such that they are all distinct. Then we decode all the test patterns with BDD, obtaining a set of candidate codewords $\vec{w}^{(m)}_{j}$, where $m\in \{1,2\}$.
	
	\begin{figure*}[tb]
		\centering
		\input{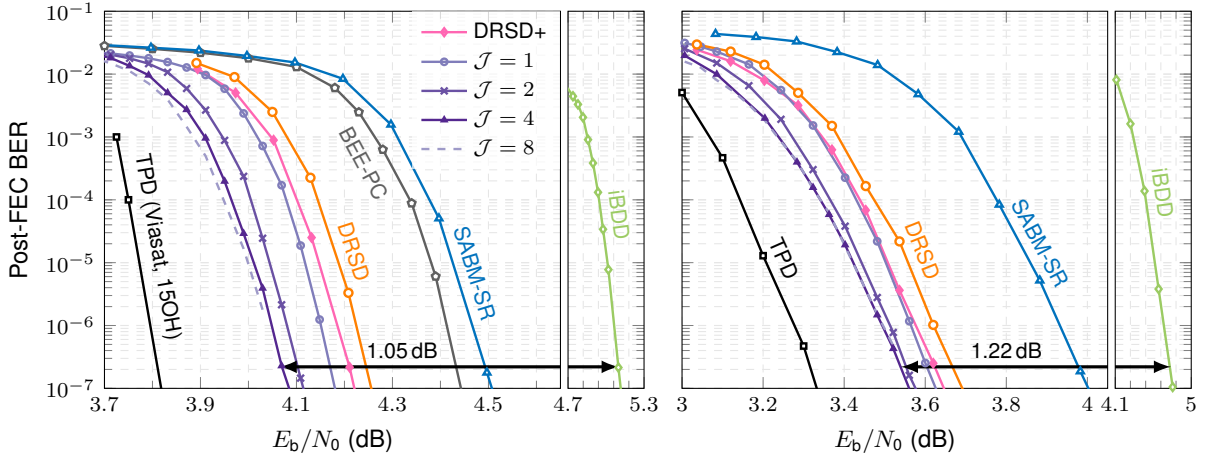}
		\vspace{-1\baselineskip}
		\caption{Post-BER vs. $\Eb/\No$ curves for PCs of rate 0.87 (15\% overhead, left plot) and rate 0.78 (28\% overhead, right plot).}
		\label{fig:ber}
	\end{figure*}
	For each obtained candidate codeword, we will first determine if it is a \ac{MC} using the DRSs. Let $\vec{w}$ be one of the candidate codewords. \ac{MC} detection is performed for $\vec{w}$ using\vspace{-0.5\baselineskip}
	\begin{equation*}
	\na = \max\{\textnormal{DRS}_i:  w_i\neq y_i,i=1,2,\ldots,n\}\vspace{-0.5\baselineskip}
	\end{equation*}
	and\vspace{-0.5\baselineskip}
	\begin{equation*}
	\nee = \sum_{w_i\neq y_i, y_i\neq ?} (\textnormal{DRS}_i+1),\vspace{-0.5\baselineskip}
	\end{equation*}
	where $i\in \{1,2,\ldots,n\}$. $\vec{w}$ is identified as miscorrection if $\na\geq \Ta$ or $\nee\geq \Te$. The first condition requires that $\vec{w}$ conflicts with bits which are considered highly reliable. The second condition requires that the sum of the DRSs of the bits that are flipped in $\vec{w}$ is large. The threshold parameters $\Ta$ and $\Te$ are calculated as described in the following. Let $\vec{r}=\left(r_1,r_2,\ldots, r_L\right)$ be vector of the mean values of the DRSs in each iteration. Define two offset vectors $\vec{a}=\left(a_1,a_2,\ldots, a_L\right)$ and $\vec{b}=\left(b_1,b_2,\ldots, b_L\right)$ associated with the thresholds. In the $\ell$-th iteration, let $\Ta=r_{\ell}+a_{\ell}$ and $\Te = r_{\ell}+b_{\ell}$. The offset vectors are optimized numerically during the decoder design. A candidate codeword $\vec{w}$ is only considered valid if it passes the miscorrection check. Then the $\DFC$ decoding result $\vec{c}$ is the one valid $\vec{w}^{(m)}_{j}$ with the smallest Hamming distance to $\vec{y}$ at the unerased positions of $\vec{y}$. In the case of multiple candidate codewords with the same distances, one of them is chosen randomly.
	
	After the $\DFC$ decoding result is determined, we update the DRSs of the bits in this row/column. If $\vec{y}\in \CW$ (indicated by a zero syndrome), no actual decoding is carried out and the DRSs for all $y_i$ are increased by one. If $\DFC$ fails to find any codeword (no candidate codeword found or all of them were identified as miscorrections), no update of the DRS is required. If $\DFC$ successfully finds a valid codeword $\vec{c}$, we reduce the DRS for all the bits which are flipped by the decoding decision (i.e., $c_i\neq y_i$) by one. The value of DRSs are clipped to $[0,31]$.
	
	We decode until the maximal number of iterations $L$ is reached or until the entire PC block is successfully decoded. Unresolved erasures are replaced by a binary random value.
	\vspace{-0.5\baselineskip}
	\section{Simulation Results}
	
	We test our proposed decoder with two exemplary PCs constructed from $[255,238,2]$ and $[127,112,2]$ component codes with PC rate $0.87$ and $0.78$, respectively. The post-FEC \ac{BER} results for the proposed decoder with different $\mathcal{J}$ values are obtained by Monte-Carlo simulations and are plotted in Fig.~\ref{fig:ber}. For reference, we compare our decoder with the decoding result of DRSD~\cite{miao2021improved} and its variant DRSD+~\cite{miao2022journal}, as well as with \ac{TPD}~\cite{pyndiah1998near}, SABM-SR~\cite{liga2019novel}, BEE-PC~\cite{sheikh2021novel} with the data points provided in~\cite{sheikh2021novel} and~\cite{liga2019novel}, and additionally the commercially-used Viasat TPD with $15$\% overhead~\cite{ViasatTPC66100}. Some of the results of the reference decoders have been evaluated using PCs with a small but negligible rate difference from ours and possibly a different maximal iteration number $L$.
	For the proposed decoders used to decode the two PCs in Fig.~\ref{fig:ber} with $L=20$, the erasure thresholds are set to be $T=0.13$ for the rate $0.87$ PC and $T=0.17$ for the rate $0.78$ PC. The optimized offset vectors are given as\vspace{-0.5\baselineskip}
	\begin{equation*}
	\begin{aligned}
	\vec{a}& =
	\left(-8,-8,-7,\cdots,-7,+1,+1,+1,+\infty,+\infty\right)\\
	\vec{b}&= 
	\left(0,0,+1,+2,\cdots,+2,+\infty,+\infty\right).
	\end{aligned}\vspace{-0.5\baselineskip}
	\end{equation*}
	
	As Fig.~\ref{fig:ber} shows, the proposed decoder provides a significant improvement compared to iBDD and has only a small gap to the TPD results. When $\mathcal{J}\leq 4$, the decoding performance improves with increasing $\mathcal{J}$ and converges for larger values of $\mathcal{J}$. For $\mathcal{J}> 8$, the decoding performance improvement is negligible and not shown. The conjectured \acp{NCG} for $\mathcal{J}=4$ are $10.75\;$dB for the rate $0.87$ PC and $11.13\;$dB for the rate $0.78$ PC.
	\vspace{-0.1\baselineskip}
	\section{Decoding Complexity}
	The major computational overhead of the proposed decoder compared to \ac{iBDD} is the increased number of BDD steps. For decoding the rate $0.87$ PC (Fig.~\ref{fig:ber} left plot), the estimated number of BDD steps in the proposed decoder at a target \ac{BER} of $10^{-5}$ is increased by a factor of $0.6$ ($\mathcal{J}=1$), $1.9$ ($\mathcal{J}=2$), $2$ ($\mathcal{J}=4$),  or $2.9$ ($\mathcal{J}=8$) compared to $10$ iterations of iBDD. For the rate $0.78$ PC, the factors are $0.65$, $1.3$, $2$, $3$, respectively. The operations required for selecting candidate codewords are simple. During iterative decoding and DRS updating, only ternary messages are passed. The additional storage space required for storing the DRSs is smaller than the space required by TPD. Our novel algorithm has a significantly lower decoding complexity than TPD.
	
	\vspace{-0.1\baselineskip}
	\section{Conclusion}
	The proposed decoder provides a good performance-complexity trade-off for decoding PCs. The high \acp{NCG} make this scheme a promising candidate for future low-complexity optical communication systems.
\printbibliography

	\vspace{-4mm}
	
\end{document}